\documentclass{rspublic}

\usepackage{graphicx}
\usepackage{epsfig}

\def\ltsima{$\; \buildrel < \over \sim \;$}
\def\simlt{\lower.5ex\hbox{\ltsima}}
\def\gtsima{$\; \buildrel > \over \sim \;$}
\def\simgt{\lower.5ex\hbox{\gtsima}}
\def\lya{Ly$\alpha$}

\begin{document}

\title[Element abundances]{The first galaxies: clues from element abundances}

\author[M. Pettini]{Max Pettini}

\affiliation{Institute of Astronomy, Cambridge, England}

\label{firstpage}

\maketitle

\begin{abstract}{}

It has recently become possible to measure directly the
abundances of several chemical elements in a variety of environments
at redshifts up to $z \simeq 5$. In this review I summarise
the latest observations of Lyman break galaxies, damped Lyman alpha
systems and the Lyman alpha forest with a view to uncovering any
clues which these data may offer to the first episodes of star
formation. The picture which is emerging is one where the
universe at $z = 3$ already included many of the components
of today's galaxies---even at these early times we see
evidence for Populations I and II stars, while the `smoking gun'
for Population III objects may be hidden in the chemical
composition of the lowest density regions of the IGM, yet to be
deciphered.
\end{abstract}

\section{Introduction}

The aim of this talk is to consider the information on the first
episodes of star formation in the universe provided by studies of
element abundances at high redshift. This is very much a growth
area at present. Thanks largely to the new opportunities offered 
by the Keck telescopes and their VLT counterparts in the southern
hemisphere, we find ourselves in the exciting position of being
able, for the first time, to measure the abundances of a wide 
range of chemical elements directly in stars, H~II regions, cool
interstellar gas and hot intergalactic medium, all observed when
the universe was only $\sim 1/10$ of its present age. Our
simple-minded hope is that, by moving back to a time when the
universe was young, clues to the nature, location, and epoch of
the first generations of stars may be easier to interpret than in
the relics left today, some 11~Gyrs later. Furthermore, as we
shall see, the metallicities of different structures in the
universe and their evolution with redshift are key factors 
to be considered in our attempts to track
the progress of galaxy formation through the cosmic
ages. As can be readily appreciated from inspection of Figure 1,
our knowledge in this field is still very sketchy. Given the
limitations of space, this review focuses primarily on
results obtained in the last year on the three components of
the high $z$ universe shown in Figure 1. 

%
%
\begin{figure} 
\vspace*{-1.5cm} 
\hspace*{-1.8cm}
\psfig{figure=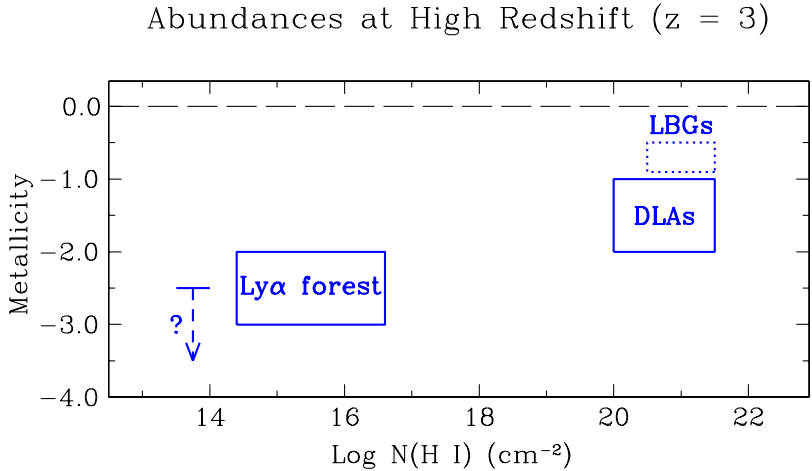,width=12cm,angle=270} 
\vspace*{-2.75cm} 
\caption[]{Summary of our current knowledge of
abundances at high redshift. Metallicity is on a log scale
relative to solar, and $N$(H~I) is the column density of neutral
hydrogen measured in the \lya\ forest, damped \lya\ systems and
Lyman break galaxies. \\ 
} 
\end{figure}

\section{Lyman Break Galaxies}
Undoubtedly, one of the turning points in extragalactic 
astronomy in the 1990s has 
been the realisation that high redshift galaxies can be
found in large numbers 
using a highly efficient photometric selection 
technique based on
the passage of the Lyman edge---at the rest wavelength of 912~\AA---through 
the $U$-band. 
After many years of fruitless searches (targeted mainly to
\lya\ emission which turned out to be a 
less reliable marker than anticipated),
we have witnessed a veritable explosion of data from the 
{\it Hubble Deep Fields} and ground-based surveys. 
Galaxies with {\it measured} redshifts 
in excess of $z \simeq 2.5$ now number in the many hundreds
(the 1000 mark is just around the corner); such large samples 
have made it possible to trace the star formation history of the 
universe over most of the Hubble time and to measure large-scale properties
of the population, most notably their clustering and luminosity 
functions (Madau {\it et al.} 1996; 
Steidel {\it et al.} 1998, 1999 and references therein).

However, for a quantitative study of many of the physical properties
of Lyman break galaxies, even the light-gathering power of the world's 
largest telescopes is not enough and we have to rely on gravitational lensing
to boost the flux to levels where their spectra can be recorded 
with sufficiently high resolution and signal-to-noise ratio.
This is the case for the $z = 2.73$ galaxy 
MS~1512-cB58, an $L^{\ast}$ Lyman break galaxy fortuitously
magnified by a factor of $\sim 30$ by the foreground cluster
MS~1512+36 at $z = 0.37$~ (Yee {\it et al.} 1996; Seitz {\it et al.} 1998).
Somewhat ironically, our Keck spectrum of cB58 (Pettini {\it et al.} 2000{\it b})
is one of the best examples
of the ultraviolet spectrum of a starburst galaxy {\it at any redshift},
thanks to the combined effects of gravitational lensing, 
redshift, and collecting area of the Keck telescopes.

%
%
\begin{figure}
\vspace*{-1.75cm}
\hspace*{-0.75cm}
\psfig{figure=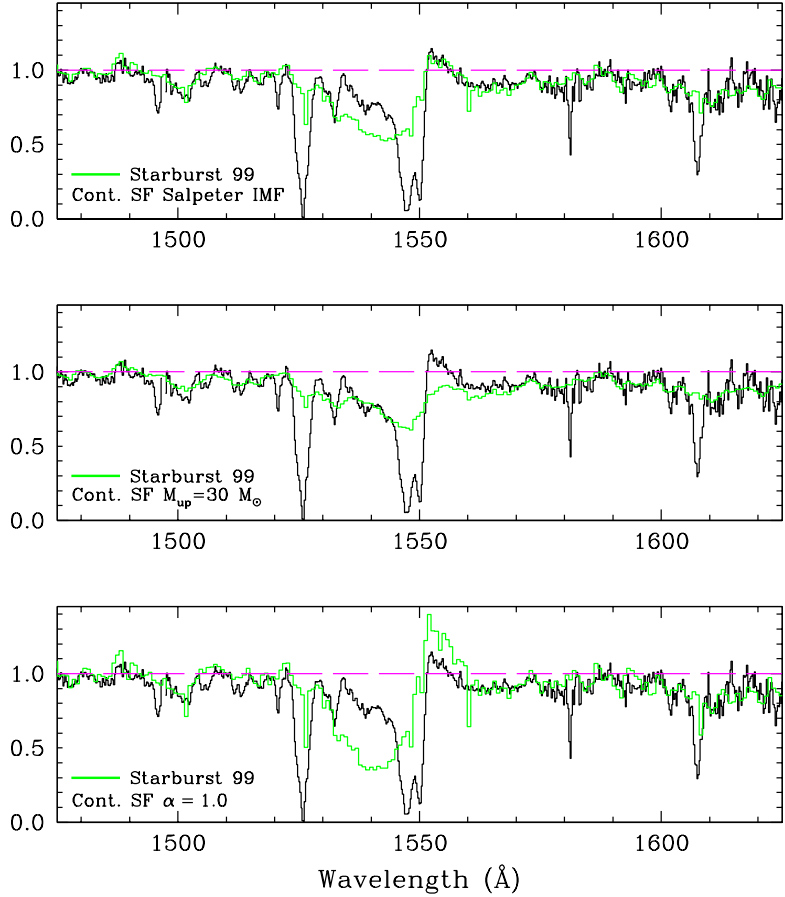,width=14cm}
\vspace*{-3.2cm}
\caption[]{Comparisons between {\it Starburst99} (Leitherer {\it et al.} 1999) 
population synthesis models with different IMFs (green lines---grey in the
black and white version of the figure)
and the Keck spectrum of MS~1512-cB58 analysed by
Pettini {\it et al.} (2000{\it b}) in the region
near the C~IV doublet (black histogram).}
\end{figure}

At $z \simeq 3$ optical wavelengths correspond to the rest-frame far-UV,
where we see the integrated light of short-lived O and early B stars.
Such spectra are most effectively analysed with 
population synthesis models, 
the most sophisticated of which is {\it Starburst99} developed by the 
Baltimore group (Leitherer {\it et al.} 1999). In Figure 2 we compare 
{\it Starburst99}\/ model predictions 
for different IMFs with our data in the region of
the C~IV~$\lambda\lambda 1548, 1550$ doublet. It is important to realise
that the comparison only refers to {\it stellar} spectral features
and does not include the {\it interstellar} lines, 
readily recognisable by 
their narrower widths (these IS lines are much stronger in cB58, where we 
sample the whole ISM of the galaxy, than in the models which are based on
libraries of nearby Galactic O and B stars). With this clarification,
it is evident from Figure 2 that the spectral properties of at least 
this Lyman break galaxy are remarkably similar to those of present-day 
starbursts---a continuous star formation model with a Salpeter IMF
provides a very good fit to the observations.
In particular, the P-Cygni profiles of C~IV, Si~IV and N~V are
sensitive to the slope and upper mass limit of the IMF; the best fit in 
cB58 is obtained with a standard Salpeter IMF with slope $\alpha = 2.35$
and $M_{\rm up} = 100 M_{\odot}$ (top panel of Figure 2). 
IMFs either lacking in the most massive stars
or, conversely, top-heavy seem to be excluded by the data
(middle and bottom panels of Figure 2 respectively).

The only significant difference between the observed and synthesised
spectrum is in the optical depth of the P-Cygni absorption trough 
which is lower than predicted (top panel of Figure 2). This is likely to
be an abundance effect, since an analogous weakening of the absorption 
is seen in OB stars with mass loss 
in the Magellanic Clouds (e.g. Lennon 1999)
and is also predicted by stellar wind theory (e.g. Kudritzki 1998). 
In future, when the libraries of stellar spectra in {\it Starburst99}
are expanded to include Magellanic Cloud stars 
(a project which is already underway),
it may be possible to calibrate the optical depth of C~IV absorption 
with Carbon abundance
and use this feature to deduce 
the metallicity of high redshift star-forming 
galaxies. For the moment, we conclude on the basis of 
a qualitative comparison that
the metallicity of the young stellar population in cB58 is similar to
that in the LMC, 
where [C/H]$ \simeq -0.6$\,.
Weak interstellar lines of Sulphur, Silicon, and Nickel 
are consistent with this abundance estimate.   

\subsection{Moving to the Infrared}
Very recently, the successful commissioning of 
NIRSPEC on Keck~II and ISAAC on VLT1 
have made it possible to extend spectroscopic
studies of Lyman break galaxies to the near infrared 
which, at $z \simeq 3$, includes the familiar optical 
emission lines from H~II regions on
which much of our knowledge of local 
star-forming galaxies is based.
As indicated by exploratory observations with UKIRT
(Pettini {\it et al.} 1998),
detecting these lines is a challenging task 
even with large telescopes, so that we 
may be restricted to studying the brightest
examples of Lyman break galaxies, with $L \simgt L^{\ast}$.
Figure 3 shows an example of such data.
The relative strengths of [O~III] and H$\beta$ in Q0201--C6
are typical of the dozen or so objects
observed so far; we find that generally 
$I_{\rm H\beta} \simlt I_{4959}$
and $R_{23} \simgt +0.7$, where
$R_{23}$ is the familiar strong line ratio index
of Pagel {\it et al.} (1979).
This in turn implies abundances of $\approx 1/3 - 1/6$ solar;
as shown by Teplitz {\it et al.} (2000), cB58  
conforms to this pattern 
with [O/H]~$\simeq -0.5$, in good 
agreement with the UV analysis discussed above.

%
%
\begin{figure}
\vspace*{-2.5cm}
\hspace*{-2.75cm}
\psfig{figure=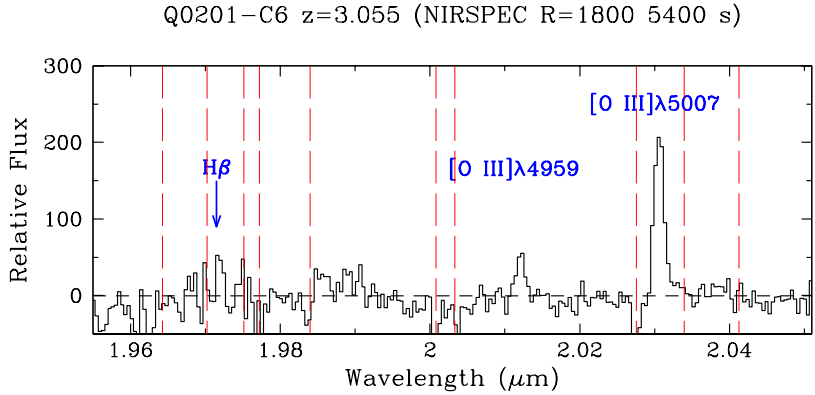,width=13cm,angle=270}
\vspace*{-3.5cm}
\caption[]{K-band spectrum of the Lyman break galaxy Q0201-C6. 
The dashed vertical lines indicate the locations of OH emission
lines from the night sky; although the lines have been subtracted out
the spectrum remains very noisy at these wavelengths. It is thus essential to
select LBGs at redshifts which place the nebular lines of interests 
in the gaps between OH emission, as is the case here.\\
}
\end{figure}

\subsection{Kinematics}
The combination of (rest-frame) optical and UV observations 
gives insights into several other properties of LBGs, 
apart from chemical abundances. 
The widths of the emission lines are likely to be better
indicators
of the underlying masses than the interstellar absorption lines which, 
being sensitive to very low column densities, can be broadened 
by gas accelerated to high velocities by supernovae and 
stellar winds associated with the 
star-formation activity. 
A preliminary analysis of the dozen objects in our sample indicates 
velocity dispersions $\sigma \simeq 60 - 120$~km~s$^{-1}$;
$\sigma \simeq 80$~km~s$^{-1}$ seems to be typical.

%
%
\begin{figure}
\vspace*{-1.5cm}
\hspace*{-2.5cm}
\psfig{figure=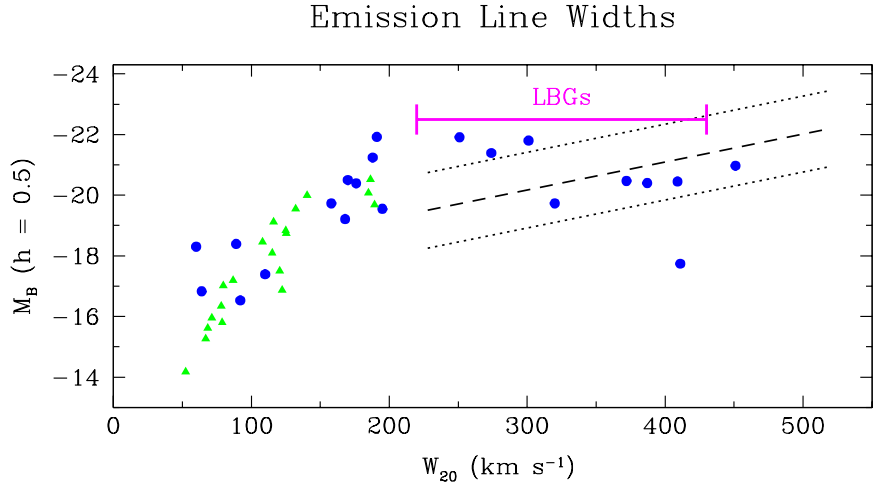,width=13cm,angle=270}
\vspace*{-3.8cm}
\caption[]{Emission line widths in Lyman break galaxies at $z \simeq 3$
and in local galaxies. 
The horizontal bar spans the range of widths at 20\% peak
intensity of [O~III] lines in a dozen Lyman break galaxies, while the 
vertical lines at each end of the 
bar indicate the range of luminosities sampled (Pettini {\it et al.} in 
preparation).
The light-coloured triangles
are the most reliable H$\beta$ measurements for H~II galaxies 
by Melnick, Terlevich, \& Terlevich (2000), 
while the filled dots are [O~II] measurements 
in a variety of nearby star forming galaxies by 
Kobulnicky \& Gebhardt (2000). Also shown is the 
relation for local spirals based on  H~I 
21cm rotation curves (broken line) and its 
$3 \sigma$ limits (dotted lines---Pierce \& Tully 1992). 
A $H_0 = 50$~km~s$^{-1}$~Mpc$^{-1}$, $\Omega = 1$
cosmology was adopted.\\
}
\end{figure}

In Figure 4 we compare these values with analogous data
for nearby galaxies. 
$W_{20}$ is full width at 20\% of the peak intensity 
($W_{20} = \sigma \times 3.62$ for a Gaussian profile)
which in the Lyman break galaxies we measure most accurately from 
[O~III]~$\lambda 5007$, and $M_B$ is the absolute magnitude in the 
rest-frame $B$-band which at $z \simeq 3$
can be deduced directly from the observed $K$-band magnitude 
without the need for a 
substantial $k$-correction.
The horizontal bar in Figure 4 
shows the range 
of values of $W_{20}$ for the LBGs observed so far, 
which are mostly 
at the bright end of the luminosity function, with 
$M_B = -22$ to $-23$ (as indicated by the vertical bars).
The most appropriate comparison is probably with the compilation of
[O~II] widths in local star-forming galaxies (filled dots) by
Kobulnicky \& Gebhardt (2000)
who mimicked the conditions under which these measurements are
conducted at high redshift, by obtaining global values of $W_{20}$ which 
refer to a {\it whole} galaxy, and did not correct
for inclination and internal 
extinction. As can be seen from Figure 4, the widths of the emission 
lines in the brighter Lyman break galaxies
are at the upper end of the range of values observed locally, and are 
significantly larger than those of H~II galaxies (triangles).
Thus the typical $\sigma \simeq 80$~km~s$^{-1}$ of LBGs
is the value which one may expect from
a disk galaxy (viewed at a random inclination) rotating at 
$\approx 150$~~km~s$^{-1}$. Indeed there are hints 
in two of the objects observed
that we may be seeing the rotation curve directly in spatially 
resolved [O~III]$\lambda 5007$ emission lines.
Kinematical masses in excess of a few times $10^{10}~M_{\odot}$ are 
indicated.

Also shown in Figure 4 is the Tully-Fisher relation for local spirals.
This comparison is less straightforward, because the Tully-Fisher
relation is derived from spatially resolved rotation curves corrected
for inclination and internal extinction---we have tried to take these
factors into account in a statistical sense when reproducing 
the mean relation by Pierce \& Tully (1992) in Figure 4.
Vogt {\it et al.} (1997) found a 
mild brightening, by $\simlt 0.4$~ mag, of the relation in a sample of 
16 galaxies at $0.15 < z < 0.75$ and 
interpreted it as being due to luminosity evolution in the 
field galaxy population
(the Vogt {\it et al.} data are, like ours, based on observations of 
[O~II] and [O~III] emission lines from H~II regions).
Taken at face value, the very preliminary comparison in Figure 4 suggests 
a much more significant luminosity evolution when we look back to 
$z \simeq 3$, perhaps amounting to as much as 2 magnitudes in 
the $B$-band.

Finally, we find that there are systematic velocity offsets
between nebular emission lines, 
interstellar absorption lines, and \lya\ in most Lyman break galaxies 
observed; these offsets can be  
explained as 
resulting from large scale outflows
with velocities of up to several hundred km~s$^{-1}$. 
In cB58 Pettini {\it et al.} (2000{\it b}) deduced a mass outflow rate
$\dot{M}  \simeq 60~M_{\odot}~{\rm yr}^{-1}$,
comparable to the rate at which gas is being turned into stars.
Such galactic `superwinds' seem to be a
common feature of starburst galaxies at all redshifts
(see Tim Heckman's article in this volume), and may well be
the mechanism which self-regulates star formation, distributes
metals over large volumes and allows the escape of ionizing photons into
the intergalactic medium.

In summary, all the available information is consistent with the 
notion that Lyman break galaxies are already well developed systems
at $z \simeq 3$, with stellar populations, chemical abundances and kinematics
very much in line with those of the more massive star-forming galaxies
in the local universe.
As explained above,
the best studied examples so far are all
at the bright end of the luminosity function; 
thus, perhaps we should not be 
surprised to find that their properties are relatively uniform.
What is interesting is 
that galaxies at such an advanced stage of evolution
were already in place at the relatively early epochs 
corresponding to
$z = 3 - 4$; it therefore seems most natural to associate 
these objects with the progenitors of today's elliptical galaxies
and bulges of spirals, as proposed by Steidel et 
al. (1996).

\section{Damped Lyman alpha Systems}
These are the absorption systems  
with the highest column density of neutral hydrogen,
$N$(H~I)$\geq 2 \times 10^{20}$~cm$^{-2}$,
seen in the spectra of QSOs and they
provide us with the best opportunity 
to measure accurately the abundances of a wide 
range of elements at high redshift. The reason is simple: 
QSOs can be several hundred times brighter than Lyman break galaxies
at the same redshift. 
Surveys with HIRES on Keck~I have produced data of 
exquisite quality---a 10\% accuracy in the determination of 
interstellar gas-phase abundances is achievable with only
modest efforts (Lu {\it et al.} 1996; Prochaska \& Wolfe 1999).

This makes it all the more frustrating that 
a clear connection between 
DLAs and galaxies has yet to be established.
In principle, selecting galaxies by their 
H~I absorption cross-section 
should provide a more representative 
sampling of the field population at a given redshift than 
conventional magnitude limited surveys, either in the continuum or 
emission lines.
Thus one may conjecture that DLAs pick out galaxies 
from the whole luminosity function, particularly if 
H~I cross section has only a mild dependence on galaxy 
luminosity (Steidel, Dickinson, \& Persson 1994), 
and that Lyman break galaxies may just be the most 
luminous DLAs at $z \simeq 3$ 
(Pettini {\it et al.} 2000{\it b}; Steidel, Pettini, \& 
Hamilton 1995; Djorgovski {\it et al.} 1996).
While this interpretation is attractive in its simplicity, 
we must face up to the fact that it cannot readily account for the 
most recent observations of DLAs 
(reproduced in Figure 5)  which find no significant 
evolution of either the gas mass or the 
metallicity (Rao \& Turnshek 2000; Pettini {\it et al.} 1999; Prochaska \& 
Wolfe 2000)
over a redshift interval ($z \simeq 0.5 - 4$)
during which most of today's stars were apparently formed
(see Figure 7 of Pettini 1999).
Possibly, existing samples of
damped \lya\ systems are subject to subtle selection effects of 
their own and may preferentially trace a particular stage in the 
evolution of galaxies, when the gas has an extended distribution 
and only moderate surface density, and the metal---and therefore 
dust---content is low. There is both theoretical 
(Mo, Mao, \& White 1999) and 
observational (Le Brun {\it et al.} 1997; Rao \& Turnshek 1998)
evidence
in support of this picture. 

%
%
%
\begin{figure}
\vspace*{-1.5cm}
\hspace*{-0.75cm}
\psfig{figure=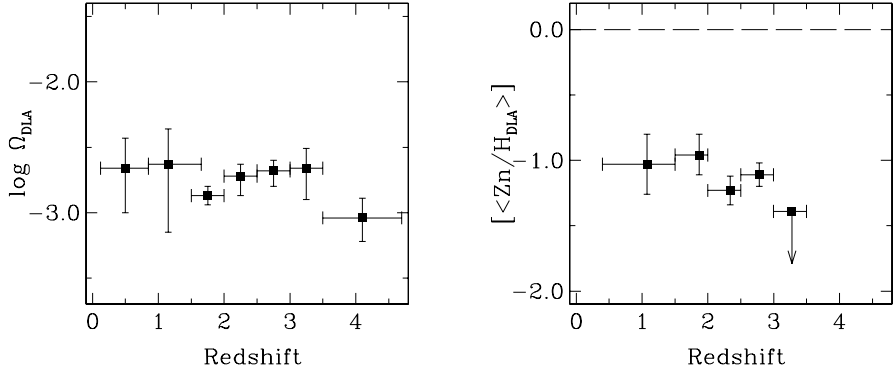,width=10.5cm,angle=270}
\vspace*{-2cm}
\caption[]{(Lack of) redshift evolution in the comoving mass density 
(left, reproduced from Rao \& Turnshek 2000; 
$H_0 = 65$~km~s$^{-1}$~Mpc$^{-1}$, $\Omega = 1$) and
metallicity (right, from Pettini {\it et al.} 1999)
of damped \lya\ systems.\\
}
\end{figure}

These latest developments do not detract from our interest in damped 
\lya\ systems. First, as emphasised repeatedly by 
Fall and collaborators (Fall 1996),
the column density-weighted mean metallicity of DLAs is
the closest measure we have of the degree of metal enrichment 
reached by the gaseous component of galaxies at a given epoch, 
{\it irrespectively of the precise nature of the absorbers}.
Thus, values of [$\langle$Z$_{\rm DLA}\rangle$]
at different redshifts
(so far most effectively deduced from the 
abundance of Zn---Pettini {\it et al.} 1999)
are essential reference points for 
models of global chemical 
evolution (Prantzos \& Silk 1998; Pei, Fall, \& Hauser 1999).
The only uncertainty which remains to be resolved for a
full use of this information is the degree to which
existing samples of DLAs are biased
against sight-lines sufficiently dusty to obscure 
the background QSOs; this is a question which we are 
in the process of exploring by examining the statistics of damped systems in
radio selected QSOs.
At redshifts $z \simeq 2 - 3$ the metallicity distribution of known DLAs
is intermediate between those of stars in the halo and thick disk of the 
Milky Way; at this epoch most of the galaxies giving rise to 
damped systems were clearly less evolved chemically 
than the stellar population forming the thin disk of our Galaxy
(see Figure 6).

%
%
\begin{figure}
\hspace{0.75cm}
\includegraphics[width=0.9\textwidth]{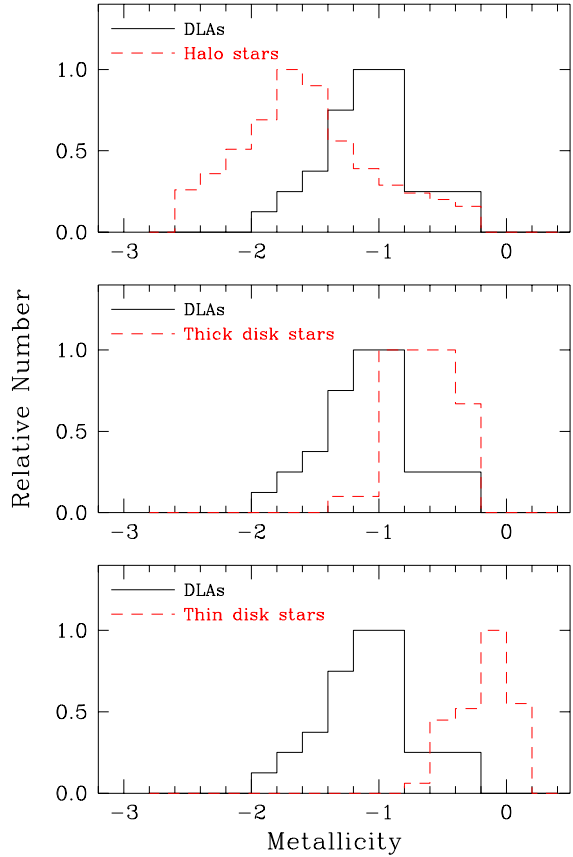}
\vspace{-1cm}
\caption[]{Metallicity distributions, normalised to unity,
of DLAs at $z \simeq 2 - 3$ 
and of stars belonging to the disk and halo populations in the 
Milky Way. See Pettini {\it et al.} (1997) 
for references to the sources of data
and further details.\\
}
\label{eps4}
\end{figure}

Second, DLAs present us 
with the opportunity to extend local studies of
the relative abundances of different elements 
to unexplored regimes and to earlier epochs. 
Potentially, DLAs have an important role to play here in 
complementing the information so far obtained from observations of
Galactic stars and nearby H~II regions
and providing fresh clues both to the origin of different 
stellar populations and to the stellar yields.

%
%
\begin{figure}
\vspace{-3.5cm}
\hspace{-1.4cm}
\includegraphics[width=1.2\textwidth]{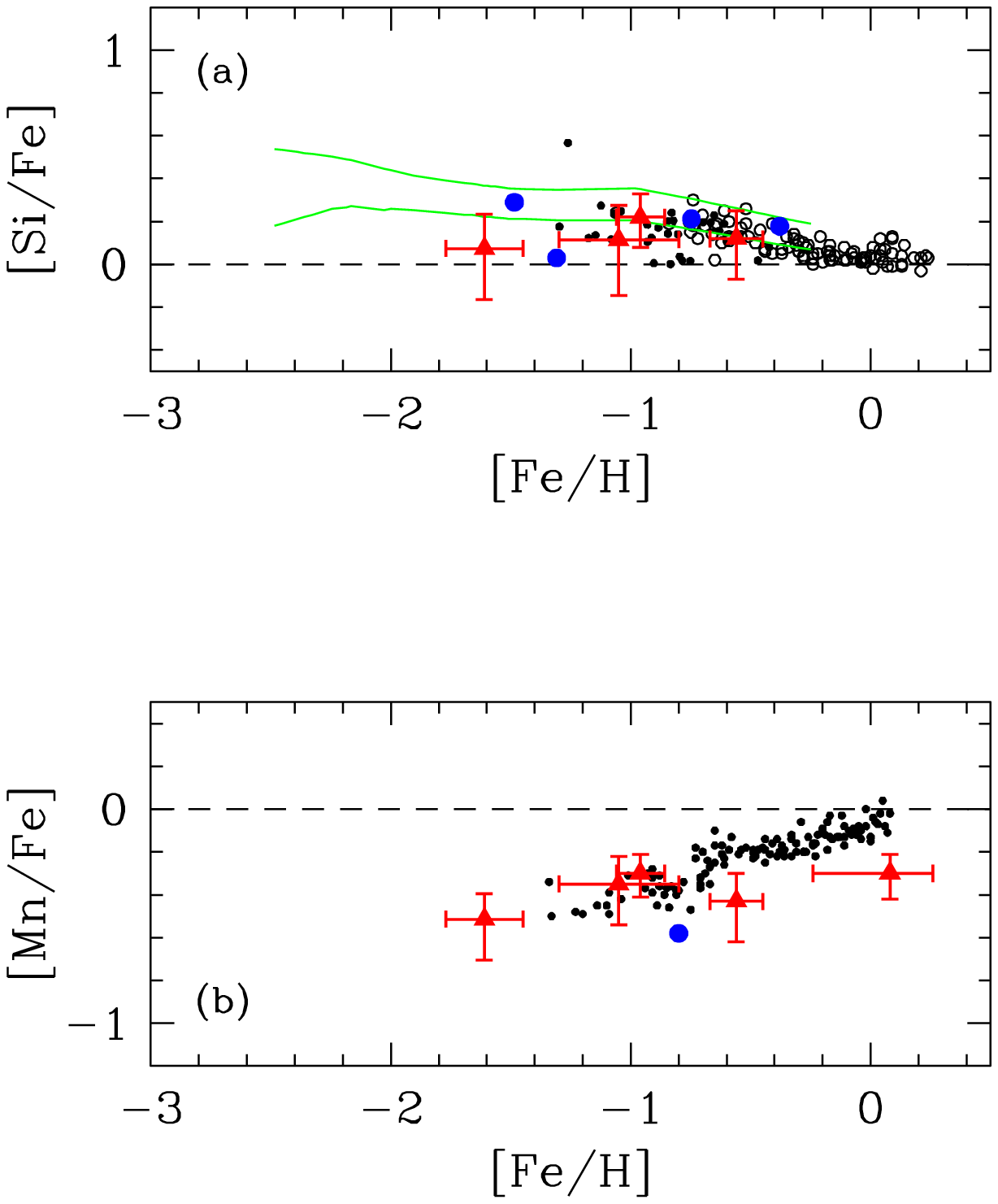}
\vspace{-3.5cm}
\caption[]{Metallicity dependence of the abundances of Si and Mn. 
Small dots are values in Galactic stars 
from Edvardsson {\it et al.} (1993) and Nissen \& Schuster (1997)
for Si, and from Nissen {\it et al.} (2000) for Mn. 
The continuous thin lines in the top panel show the range
(upper and lower quartiles) spanned by the compilation of measurements
in halo stars by Ryan, Norris, \& Beers (1996).
The other symbols refer to damped \lya\ systems.
}
\label{eps7}
\end{figure}

For example, 
Pettini, Lipman, \& Hunstead (1995) 
and Lu, Sargent, \& Barlow (1998)
showed that in DLAs it is possible to follow the behaviour 
of the (N/O) ratio to lower metallicities than those probed up to now
(IZw18 still remains the most metal-poor star-forming region known 
in our vicinity). Their results appear to lend support
to the idea of a delayed production of primary nitrogen by intermediate 
mass stars, although this interpretation has been challenged more recently 
(Centuri\'{o}n {\it et al.} 1998; Izotov \& Thuan 1999---see 
also Pilyugin 1999)
and more observations are clearly required in order to settle the issue.

The latest application of this technique 
involves Silicon (an $\alpha$-capture element) and Manganese.
In Figure 7 (reproduced from Pettini {\it et al.} 2000{\it a})
the abundances of these two elements 
in damped \lya\ systems of different metallicities
are compared with 
analogous data for stars in the 
disk and halo of our Galaxy.
The DLAs considered are those where less than 50\% of 
Si, Mn, and Fe is locked up in dust grains, 
so that the total (gas+dust) abundances 
can be recovered with minimum uncertainty.
The first-order conclusion is that the DLA values
roughly follow the local trends, but there are notable differences too, 
as we now discuss.

The rise of [Si/Fe] from to solar to between 
$+0.3$ and $+0.4$ as the metallicity
drops to [Fe/H]$ = -2$ (top panel of Figure 7) is the well-known 
overabundance of the $\alpha$-elements commonly attributed to the delayed 
production of additional Fe by Type Ia supernovae.
While some DLAs do show enhanced [Si/Fe], we also find counter-examples
of near-solar abundance of Si at metallicities in the range
[Fe/H]$\simeq -1$ to $-2$. 
Current wisdom would interpret such cases as arising in galaxies 
where star formation has proceeded slowly, or in bursts, so that there 
has been sufficient time for Fe to build up to solar
abundance relative to Si, while the overall metallicity 
remained low.  Corroborating evidence for 
this interpretation may be provided by deep imaging of the absorbers,
if they are found to be low surface brightness or dwarf galaxies.

Turning to Mn (lower panel of Figure 7), 
the strong decrease in [Mn/Fe] towards low metallicities 
is now well documented, but its origin is unclear. Two 
possibilities have been proposed and both have problems with the DLA 
data, at least at face value. 
If the stellar trend is due to a metallicity
dependent yield of Mn in massive stars, it is difficult
to explain the one DLA with [Mn/Fe] $\simeq -0.3$ at solar [Fe/H].
On the other hand, enhanced (relative to Fe)
production of Mn in Type Ia 
supernovae cannot explain DLAs with low [Mn/Fe] and near-solar
[Si/Fe] of which there are at least two examples.

In concluding this section, it is important to stress the preliminary 
nature of the above conclusions which are based on the comparison of very 
few measurements in DLAs with a much larger body of stellar data. 
One of the lessons from stellar work is that there is considerable 
scatter, both observational and intrinsic,
in the relative abundances of different elements so that most trends only 
become apparent when a large set of measurements has been assembled.
Thus Figure 7 should be taken as no more than 
an illustration of the issues which 
can be addressed with surveys of element abundances in damped 
systems. Although work on element ratios 
in high redshift galaxies is still a long way behind 
its counterpart in Galactic stars,
it may well 
play a decisive role in resolving some outstanding questions on
the origin of elements.

\section{The Lyman alpha Forest}

The final component of the high redshift universe considered in 
this review is the all-pervading intergalactic medium which manifests 
itself as fluctuating absorption bluewards of the \lya\ emission line
of every QSO. 
Observationally, the term \lya\ forest is used to indicate the bulk of 
discrete \lya\ absorption lines with column densities in the range
$10^{16} \simgt N{\rm (H I)} \simgt 10^{12}$~cm$^{-2}$; 
since this gas is 
highly ionised, it may account for most of the baryons at 
$z \simeq 3$ (Rauch 1998).
Hydrodynamical simulations have shown that the \lya\ forest 
is a natural consequence of structure formation in a 
universe dominated by cold dark matter and bathed in a diffuse ionising 
background (e.g. Weinberg, Katz, \& Hernquist 1998).
In this picture, the physics of the absorbing gas is relatively simple 
and the run of optical depth $\tau$(\lya) with redshift 
can be thought of as a `map'
of the density structure of the IGM along a given line of sight.
At low densities, where 
the temperature of the gas is determined by the balance between 
photoionisation 
heating and adiabatic cooling, 
$\tau$(\lya)$ \propto {\rm (}1 + \delta{\rm )}^{1.5}$, where the 
$\delta$ is the overdensity of baryons
$\delta \equiv {\rm (}\rho_{\rm b}/\langle \rho_{\rm b} \rangle - 1{\rm )}$.
At $z = 3$ $\tau$(\lya)$ = 1$ corresponds to a region of the IGM which 
is just above the average density of the universe at that time
($\delta \approx 0.6$).

The lack of associated metal lines was originally
one of the defining characteristic of the \lya\ 
forest and was interpreted as evidence 
for a primordial origin of the clouds (Sargent {\it et al.} 1980).
As it is often the case, subsequent improvements in the 
observations have shown 
this to be an oversimplification and in reality weak metal absorption,
principally by C~IV, is present at the redshift
of most \lya\ 
clouds down to the detection limit of the data (Songaila \& Cowie 1996).
The degree of metal enrichment implied is relatively high,
([C/H]$\simeq -2.5$ with a scatter of perhaps a factor of
$\sim 3$---Dav\'{e} {\it et al.} 1998), 
in the sense that stars with significantly lower 
metallicities are known to exist in the halo of our Galaxy.

It is not easy to understand how the low density IGM came to be polluted 
so uniformly by the products of stellar nucleosynthesis at such an early 
epoch.
While, as explained above,
we see directly the outflow of metal-enriched gas 
in `superwinds' from Lyman break galaxies at the same redshift,
most of this gas is not expected to travel far from the production sites,
because it is either trapped by the gravitational potential of the 
galaxies, if they are sufficiently massive, or is confined by the 
pressure of the hot IGM (Ferrara, Pettini, \& Shchekinov 2000).
Whether an early episode of pre-galactic star formation 
is required depends on whether C~IV lines continue to be seen
in \lya\ clouds of diminishing H~I column density.
Current limits are for $N$(H~I)$\simgt 3 \times 10^{14}$~cm$^{-2}$
(some 75\% of such \lya\ clouds have associated 
C~IV absorption---Songaila \& Cowie 1996)
corresponding to moderately overdense gas 
($\delta \simgt 10$)
which in the simulations is preferentially found in the vicinity of 
collapsing structures and may thus reflect local, rather than universal, 
metal pollution.

%
%
\begin{figure}
\centering
\vspace{-0.5cm}
\hspace{-0.2cm}
\includegraphics[width=.60\textwidth,angle=270]{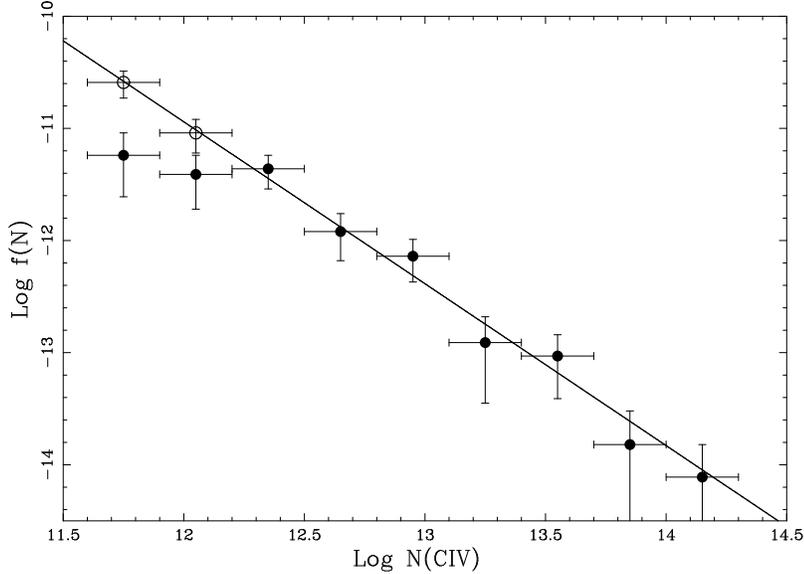}
\vspace{0.3cm}
\caption[]{ 
C~IV column density distribution in Q1422+231 
at $\langle z \rangle \simeq 3.15$
(reproduced from Ellison {\it et al.} 2000);
$f$($N$) is the number of C~IV systems per column density 
interval and per unit redshift path. 
The filled circles are the data  grouped into bins 
of 0.3 in log~$N$(C IV) for display purposes.
The line shows the best fitting power-law slope 
$\alpha = 1.44$, assuming the distribution to be of
the form $f{\rm (}N{\rm )} dN = B N^{-\alpha} dN$.
The open circles show the values corrected for incompleteness
at the low column density end; the correction factors were estimated with 
the aid of simulations which showed that only $\sim 43$\%
and $\sim 23$\% of C~IV lines with log~$N$(C IV) = 12.05 and 11.75 
respectively
are typically recovered from the data.
Once these correction factors are introduced, there is no indication of a 
turnover in the column density distribution down to the lowest values
of $N$(C IV) reached up to now.}
\label{eps8}
\end{figure}

The detection of C~IV lines when $N$(H~I)$ \simlt 1 \times 10^{14}$
is a challenging task, even with a 10-m telescope, because we are dealing 
with observed equivalent widths 
$W_{\lambda}{\rm (}1550{\rm )} \simlt 2.5$~m\AA.
A possible approach in these circumstances 
is to try and recover such a weak signal from a statistical treatment
of many lines which individually are below the detection limit. 
Unfortunately, different analyses have reached conflicting 
conclusions (Lu {\it et al.} 1998; Cowie \& Songaila 1998). 
Furthermore, a recent reappraisal of the
techniques with the help of extensive simulations 
of the spectra has indicated that
many subtle effects, such as small random differences between
the redshifts of \lya\ and C~IV absorption, 
make the interpretation of the results far from 
straightforward (Ellison {\it et al.} 1999).

A more direct way to tackle the problem is 
to push the detection 
limit further by securing spectra of 
exceptionally high signal-to-noise ratio;
as for Lyman break galaxies this is most effectively achieved
with the aid of gravitational lensing.
In this way Ellison {\it et al.} (2000) were able to
reach S/N$ = 200 - 300$ in the C~IV region between $z = 2.91$ and 3.54
of the gravitationally lensed QSO Q1422+231
after adding together data recorded over several nights 
with HIRES on Keck~I.

As can be seen from Figure 8, the number of weak C~IV lines continues to 
rise as the signal-to-noise ratio of the spectra increases and any 
levelling off in the column density distribution presumably
occurs at $N$(C IV)$ < 5 \times 10^{11}$~cm$^{-2}$. This limit is one 
order of magnitude more sensitive than those reached 
previously.
In other words, we have yet to find any evidence in the \lya\ forest for 
regions of the IGM which are truly of 
primordial composition or have abundances as low as those of
the most metal-poor 
stars in the Milky Way halo.
Pushing the sensitivity of this search even further
is certainly one of the goals for the future.

\section{Conclusions}
This review has covered a lot of ground reflecting the fast pace of
progress in the study  of element abundances at high redshift, now set to 
accelerate further with the forthcoming availability of new, efficient
spectrographs on Keck, VLT, Subaru and Gemini.
The picture which is emerging is that 
of a universe at $z \simeq 3$ with many of
today's characteristics already in place. 
At this epoch, Lyman break galaxies resembled closely 
today's star-forming galaxies   
dominated by Population~I stars,
damped \lya\ systems exhibited 
mostly Population~II chemical abundances, and the 
low density \lya\ forest may well have been the repository 
of the first heavy elements synthesised by Population~III stars.
This does not necessarily imply a 
one-to-one correspondence between these objects then and now, 
given the substantial time interval available for evolution.
It is likely that Lyman break galaxies, which at $z \simeq 3$
trace the highest peaks in the underlying mass distribution,
have through subsequent mergers
evolved into today's massive ellipticals and bulges
to be found preferentially in rich clusters.
It is also plausible that the gas giving rise to 
at least some high redshift damped \lya\ systems 
has turned into the stars of today's spiral galaxies.
And the heavy elements ejected into the IGM by the 
first stars which formed in low mass collapsed structures
have by now presumably been
augmented by the much more substantial 
metal-enriched outflows from successive generations of stars
in more massive galaxies.
Thus the lack of a clear age-metallicity relationship in our own
Galaxy is reflected on a much larger scale by the universe as a
whole---old (and high redshift) do not necessarily mean metal-poor.

What is clear is that much work still 
needs to be done before we have a full picture
of the chemical enrichment of the universe at $z \simeq 3$.
The very substantial gaps in our knowledge
evident in Figure 1 are reflected by the results of a simple
accounting exercise. 
As discussed by Pettini (1999) and more recently Pagel (2000),
the comoving density of metals so far detected in the 
\lya\ forest, damped \lya\ systems and Lyman break galaxies
accounts for only about 10\% of the total metal production 
associated with the
star formation activity we see {\it directly} at $z \simgt 3$\,.
Presumably then, as now, at least some if not most  
of the `missing metals'
are to be found in hot gas---in galactic halos and 
(proto)clusters---which has not yet been fully
accounted for, mainly because we remain ignorant of both its
metallicity and baryon content.
Food for thought, as we enter the new millennium.\\

I should like to acknowledge my collaborators in the 
various projects described in this talk, 
particularly Chuck Steidel, Sara 
Ellison, Kurt Adelberger, David Bowen, Len Cowie, Jean-Gabriel Cuby, 
Mark Dickinson, Mauro Giavalisco, Alan Moorwood, 
Joop Schaye, Alice Shapley and Toni Songaila.
I am grateful to the Royal Society and the organisers 
for inviting me to take part in such 
a stimulating discussion meeting.

\end{document}